\documentclass[preprints,review,accept,moreauthors,pdftex]{mdpi} 

\firstpage{1} 
\pubvolume{1}
\issuenum{1}
\articlenumber{0}
\pubyear{2022}
\copyrightyear{2022}
\datereceived{} 
\dateaccepted{} 
\datepublished{} 
\hreflink{https://doi.org/} 

\Title{Ionizing spotlight of active galactic nucleus  }

\TitleCitation{Ionizing spotlight of AGN}

\newcommand{\SII}{[S~{\sc ii}]}
\newcommand{\CII}{[C~{\sc ii}]}
\newcommand{\OIII}{[O~{\sc iii}]}
\newcommand{\NII}{[N~{\sc ii}]}
\newcommand{\HII}{H~{\sc ii}}
\newcommand{\HI}{H~{\sc i}}
\newcommand{\Ha}{H$\alpha$}

\newcommand{\Hb}{H$\beta$}
\newcommand{\HeII}{He\,{II}}
\newcommand{\kms}{\,\mbox{km}\,\mbox{s}^{-1}}

\newcommand{\apj}{ApJ}
\newcommand{\apjl}{ApJL}
\newcommand{\aap}{A\&A}

\newcommand{\aj}{AJ}
\newcommand{\mnras}{MNRAS}
\newcommand{\nat}{Nature}
\newcommand{\pasp}{PASP}

\Author{Alexei V. Moiseev $^{1,\dagger,\ddagger}$*\orcidA{} and Aleksandrina  A.  Smirnova }

\AuthorNames{Alexei V. Moiseev  and Aleksandrina  A. Smirnova }

\AuthorCitation{Alexei   Moiseev  and Aleksandrina   Smirnova}

\address{%
$^{1}$ \quad Special Astrophysical Observatory, Russian Academy of Sciences, Nizhny Arkhyz 369167, Russia; moisav@sao.ru
}

\corres{Correspondence: moisav@sao.ru;  (A.M.),  ssmirnova@gmail.com;  (A.S.)}

\abstract{ 
Ionization cones and relativistic jets give us one of the most large-scale example of active galactic nuclei (AGN) influence on the surrounding gas environment in galaxies and beyond. The study of ionization cones makes it possible not only to test the predictions of the unified model of galactic activity,  but also to probe galaxy gas environment and trace how the luminosity of the nucleus changes over time (a light echo). 
In the external galactic or even extragalactic gas  ionization cones create  Extended Emission-Line Regions (EELRs) which can span distances from several  to hundreds kpc away a host galaxy. 
We review the recent results of   studying the gas kinematics and  its ionization properties in EELRs with a special attention to search of fading  AGN radiation on the time scale  $\mbox{few}\times(10^4-10^5)$ yr. The role of  modern narrow-band and integral-field surveys in these researches  is also considered.
}


\begin{document}

\section{Introduction} 
\label{sec:intro}

\begin{figure*}
  \centering
\includegraphics[width=0.99\textwidth]{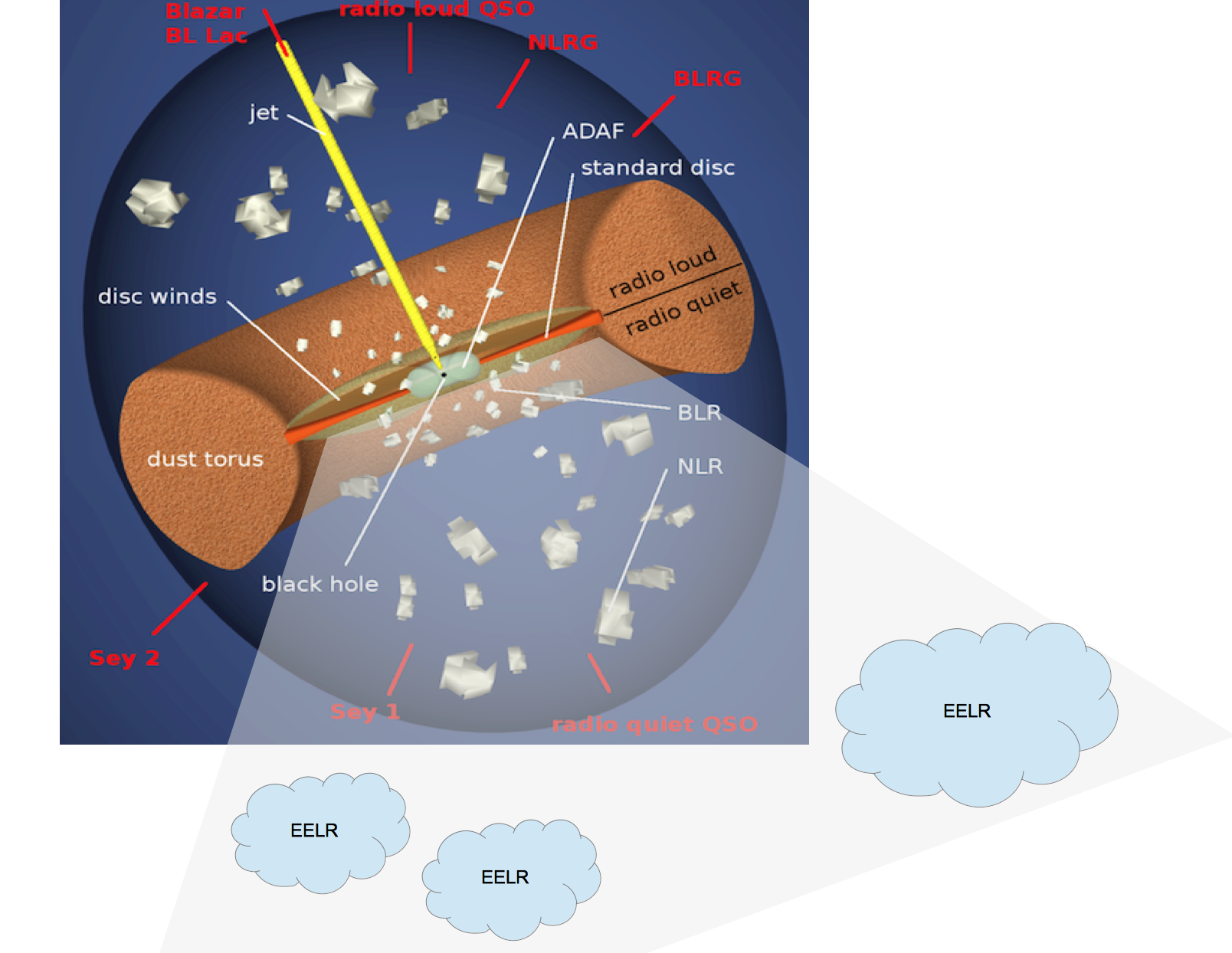}
\caption{The AGN unified model  according \citet{Brinkmann2009PhDT.......361B}.  The bottom part (the case of radio-quiet AGNs) shows the ionization cone illuminated  the external gas clouds (EELRs)
}
\label{fig1}
   \end{figure*}

There is no doubt that the  phenomenon of active galaxy nucleus (AGN) is caused by the accretion of the surrounding matter to the central super massive black hole (SMBH). Until recently, there were only indirect evidences, but now we have images of black holes in the nucleus of the active galaxy M87 \cite{EHT2019ApJ...875L...4E} and in the Milky Way \cite{EHT2022ApJ...930L..12E}, obtained by radio interferometry. Figure ~\ref{fig1} shows the  unified model of the `accretion engine' of the AGN\cite{Antonucci1993ARA&A..31..473A,Urry1995PASP..107..803U}, including a rotating accretion disk of the matter captured by the central black hole, relativistic jet and a gas-dust optically thick torus surrounding the accretion disk. The variety of observed AGN types can be explained by the different orientation  of the AGN central engine  toward the observer, so that the dust torus hides the accretion disk and the ionized gas clouds closest to the nucleus  ($r<1$ pc) in different ways. 
These clouds, moving at speeds up to several tens of thousands $\kms$, are responsible for the formation of broad components in the lines of ionized hydrogen in the spectra of type 1 Seyfert galaxy  (broad line region --- BLR). If BLR region is covered from us by a torus, the  spectrum of type 2 Seyfert galaxy is directly observed. It contains relatively narrow both  recombination and forbidden lines produced by  gas clouds at  larger distances from SMBH (narrow line region --- NLR).  
The line-fo-sight velocities do not exceed several hundreds of $\kms$, the size of a typical NLR region ranges from a few hundred parsecs to a few kpc.

\begin{figure}[]
  \centering
\includegraphics[width=0.99\textwidth]{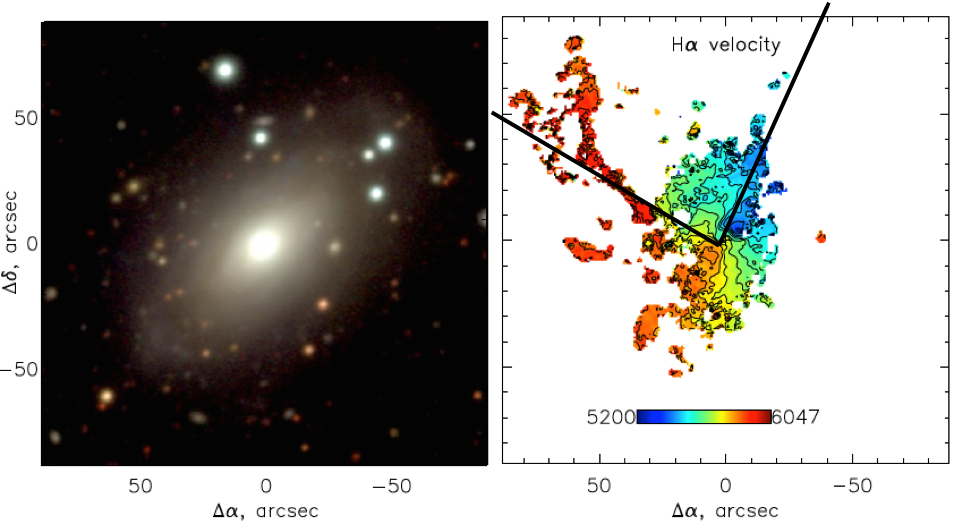}
\caption{
Observations of  Mrk~6 at the SAO RAS 6m telescope: the direct image in  broad band filters (left) and the \Ha{} velocity field  (right) according \citet{Smirnova2018MNRAS.481.4542S}. Black lines mark the ionization cone's borders according previous estimations in \citet{Kukula1996}
}
\label{Moiseev:mrk6}
\end{figure}   

The growth and activity of the SMBH is closely related to the host galaxy properties and evolution. The galaxies interaction and secular evolution \cite{Kormendy2004ARA&A..42..603K}  leads to the loss of angular momentum of the external  gas feeding  the accretion engine. On the other hand, the active nucleus  also {demonstrates}  a noticeable influence  on the surrounding interstellar medium and even intergalactic environment. Gas in accretion disk is heated up to high temperatures and radiates both as in the UV as in X-rays. Therefore, we distinguish between mechanical effects (an injection of the kinetic energy of a jet and/or hot winds to the environment) and radiative ones, associated with both ionization of the interstellar medium by energetic quanta, and with the radiation pressur from the accretion engine.  Such effect can stop star formation in the galactic gas by ionizing it and heating it up to high temperatures, so that not only fragmentation, but also the existence of cold molecular clouds becomes impossible (negative feedback). On the other hand, the  wind from outflow and shock waves associated with jet may compress the gas and trigger the star formation (positive feedback). This symbiosis between the AGN and the parent galaxy makes the activity of the nuclei an important factor in cosmological evolution \cite{Silk2012RAA....12..917S,Kormendy2013ARA&A..51..511K,IllustrisTNG2018MNRAS.473.4077P}.

The dust torus collimates the UV radiation along the disk axis in the form of two symmetric ionization cones \cite{Wilson1994AJ....107.1227W}.  In this way, apart from the BLR and NLR  the `ionization trace' of the active nucleus  may also be observed at much larger distances from the  galaxy center, depending on the presence of neutral gas both in the galactic disk itself and outside it. In the case of gas-rich disk  a classical biconical morphology in  emission lines distribution will be appeared (the traditional indicator --- the doublet of \OIII$\lambda\lambda4959,5007$). Otherwise, individual  gas clouds can be ionized both inside and outside the galactic disk (see the cone in Fig.~\ref{fig1}). The typical spatial scale can reaches  tens of kpc. Earlier, for such clouds the term ENLR (Extended NLR) was used, but now we considered  more general case, including also turbulent gas ejected from the nucleus  --- EELR (Extended emission line region, \cite{Stockton2006NewAR..50..694S, Keel2012}).

The operation of the accretion engine differs in radio-loud and radio-quiet AGN. In the first case (the upper section of the scheme in Fig.~\ref{fig1}) most of the energy goes away to the acceleration of relativistic jets particles with the subsequent emission of synchrotron radiation --- advection dominated accretion flows (ADAF). In radio-quiet AGN,   the contribution of the jets to the total luminosity is insignificant  because the most of produced radiation is associated with the accretion disk. This is characteristic of number of quasars, as well as Seyfert and LINER galaxies. These objects  make up the vast majority among nearby active galaxies and are the subject of this review. 

The ionized cone could be considered as an `ionization searchlight' that illuminates (ionizes) a spatial screen --- gas clouds. The study of such systems allows us to better understand the properties as the searchlight as the screen. We highlight the following main tasks:

\begin{itemize}
     \item Study of the gas environment of galaxies (accretion of gas clouds, cosmological filaments, etc.)
 \item `Aracheology of active galactic nuclei'  \cite{Morganti2017} --- a history of supermassive black hole activity at the light-travel time from the AGN to the gas clouds.
  \item Constrain of the unified model parameters related with the  ionization cones: measuring the cone angle, looking for traces of their precession, constraints of  absorbing matter distribution models.    
 \end{itemize}

Below we will consider  examples of solutions to each of the listed  problems, including observations performed by us and our colleagues at the Special Astrophysical Observatory of the Russian Academy of Sciences (SAO RAS) and at the Caucasus Mountain Observatory of the Sternberg  Astronomical Institute of the Lomonosov Moscow State University (CMO SAI MSU).  

\section{Filaments of the intergalactic medium} 

\subsection{The nearest example: Mrk 6}
\label{sec:Mrk6}
The   lenticular Seyfert galaxy Mrk~6 reveals one of the most nearest and impressive cases of the intergalactic environment illumination by the AGN. Mrk~6 --- one of the first known Seyfert galaxies, so it seems to have been studied in great detail \citep{Capetti1995ApJ...454L..85C,Afanasiev2014MNRAS.440..519A}:  its stellar disk looks homogeneous and symmetrical. It was also known about large (up to 22 kpc) ionization cone, aligned  with the central radio jet \cite{Kukula1996}. However  deep 3D-spectroscopic observations at  the SAO RAS 6m  telescope with the scanning Fabry-Perot interferometer \footnote{\url{https://www.sao.ru/hq/lsfvo/devices/scorpio-2/ifp_eng.html}}
revealed a  very  extended system of emission filaments  up to distances about 40 kpc (over $4$ optical radii) from the nucleus (Fig.~\ref{Moiseev:mrk6}). The velocity field in the \Ha{} emission line  showes that in the inner regions the rotation of ionized gas takes place in the plane of the stellar disk. The differences in the radial velocities of the gas from the circular rotation model here do not exceed $50\,\kms$, but outside the stellar disk the residual velocities reach $250\,\kms$. The gas is dynamically cold, i.e., it has a low velocity dispersion ($\sigma<50\kms$) and cannot be associated with jet-driven outflow \citep{Smirnova2018MNRAS.481.4542S}.

In the paper \citet{Smirnova2018MNRAS.481.4542S} we explained the line-of-sight velocity distribution by assuming that the gas in the outer emission structures rotates in circular orbits, approximately perpendicular to the host galaxy disk. The long-slit spectra of the external gas  obtained at the 6m telescope demonstrated the ionization by AGN  similar with  the inner regions of the galaxy. All available data can be explained in the assumption of ionization by AGN radiation of the gas accreted  by Mrk 6 from the intergalactic environment. Deep images in broad filters indicate the absence of any stellar structures  associated with this gas (tidal tails,   disrupted remnants of  companions, etc.). It is possible that in the ionized gas we detected part of a much larger \HI{} diffuse structure. The process of external gas accretion   is considered as an important stage of galactic  baryonic mass assembly \citep[see][for review]{accretion2014A&ARv..22...71S}.

\subsection{Cosmological filaments}

The accretion of a cold gas from large-scale structure filaments is necessary to form galaxy disk in the frameworks of   the $\Lambda$CDM cosmology, because it is a way to form  disk with a  realistic angular momentum \citep[see][for review]{Silchenko2022}.  Therefore the search for fossil traces of this process in the Local Universe  is actual. However  detailed mapping of  neutral hydrogen at high redshifts by modern telescopes is still difficult. Indeed,  at redshifts $z>4$ the 21 cm  line shifts to the meter wavelength, so the angular resolution decreases dramatically. However the filaments inside  ionization cone  can be studied in the optical range. A characteristic example is  the detection of a gas filament emitting in the Ly$\alpha$ line, about 290 kpc in size, near the quasar at $z=2.3$ \cite{Cantalupo2014Natur.506...63C}.   In recent years, more and more interesting results in the study of the large-scale gas distribution at  $z\approx3$ have been obtained using the   MUSE/VLT integral-field spectrograph  \citep[for example,][]{MUSE2021ApJ...923..252S,MUSE2023arXiv230915144W}.
At  this redshift the Ly$\alpha$ resonance line falls into the optical range. Results for more distant gas structures in the ionization cones of the Early Univers   AGN should be expected in the  observations of James Webb Space Telescope (Sec.~\ref{sec:outflow}).

\section{Galaxies with fading activity}

\subsection{Duty cycle of AGN}

The well-known tight relations between mass of SMBH and properties of its host galaxy  (total mass of the spherical component, galaxy luminosity, etc. \cite{Gebhardt2000ApJ...539L..13G,Ferrarese2000ApJ...539L...9F, Kormendy2013ARA&A..51..511K}) implies that the most disk and elliptical galaxies  harbor a sufficiently massive black hole. However  we observe the AGN phenomenon only in  a small fraction  of galaxies. From the simplest statistical considerations it's clear that duty cycle of   activity is a  relatively short  phase of the galactic nucleus life. Indeed,    galactic disk matter needs to lost  its angular moment significantly \citep[in $\sim10^4$ times, according][]{Jogee2006LNP...693..143J} to move from radius of a few kpc to the nuclear region  for consumption by a SMBH. On the one hand, we know a number of processes leading to the angular momentum lost and transfer a matter from several kpc to hundred pc radii orbits: tidal interactions, bars, circumnuclear  spirals, binary SMBHs, etc. \cite{Combes2001sac..conf..223C, Jogee2006LNP...693..143J}. On the other hand, it is often unclear what feeding mechanism is released  in every galaxy, because the timescale of  AGN duty cycle differs from characteristic dynamical time of  large-scale processes like a galactic merging. The observational information about  angular momentum transfer inside inner  10-50 pc is also insufficient, however new ALMA observations shed some light in the case of some nearby Seyfert galaxies \citep{Combes2021IAUS..359..312C}. Moreover, the negative feedback effects  caused by AGN radiation and/or outflow can terminate the processes of  gas inflow. The duration of the active phase and its recurrence is the subject of  an intense discussion \cite{Morganti2017}. Therefore, it is important and interesting to catch the galactic nucleus at the moment when the accretion machine is switch on/off.  

Spectral  monitoring of nearby AGN and  wide-field spectral  surveys  have revealed a population of galaxies where  broad emission lines suddenly appeared or, conversely, completely disappeared  (CL-AGN --- Changing look AGN). Such event may indicate a change in the geometry of the dusty region or in the accretion rate or some additional effects or their combination \cite[][see also Ili\'c et al in this proceeding]{LaMassa2015ApJ...800..144L,MacLeod2016MNRAS.457..389M, Changing-look2022arXiv221105132R, Popovic2023A&A...675A.178P}. However, the  CL-phenomenon is observed  on timescales  of several years and couldn't be connected with a global activity trend. On the other hand, it is possible to detect relic extended spatial  structures related with previous  episode of on-going activity. Low frequency radio observations reveal an example of giant (320 kpc in a projected linear size)   diffuse low surface brightness  source B2 0924+30 around elliptical galaxy which activity ceased $~\sim50$ Myr ago \citep{Shulevski2017A&A...600A..65S}. The images of some radio galaxies exhibit  possible signatures of two or even three events of activity, like in   
B0925+420 with a three pairs of  radio lobes with ages 0.4--270 Myr  \cite{Brocksopp2007MNRAS.382.1019B} {or in J1225+4011 with a similar structures correspond to ages 2, 19, and 220 Myr \citep{Chavan2023MNRAS.525L..87C}. The current progress in the study of relic  structures  around  radio-loud AGN related with development of modern  low-frequency telescopes like LOFAR was recently reviewed by \citet{Mahatma2023Galax..11...74M}.} In optical domain a study of the nucleus ionization trace in the surrounding gas allows us to detect  AGNs faded at the characteristic time scale $\sim0.1$ Myr as it considered below.

\subsection{Hanny’s Voorwerp}

\begin{figure}
\center{
\includegraphics[width=0.99\textwidth]{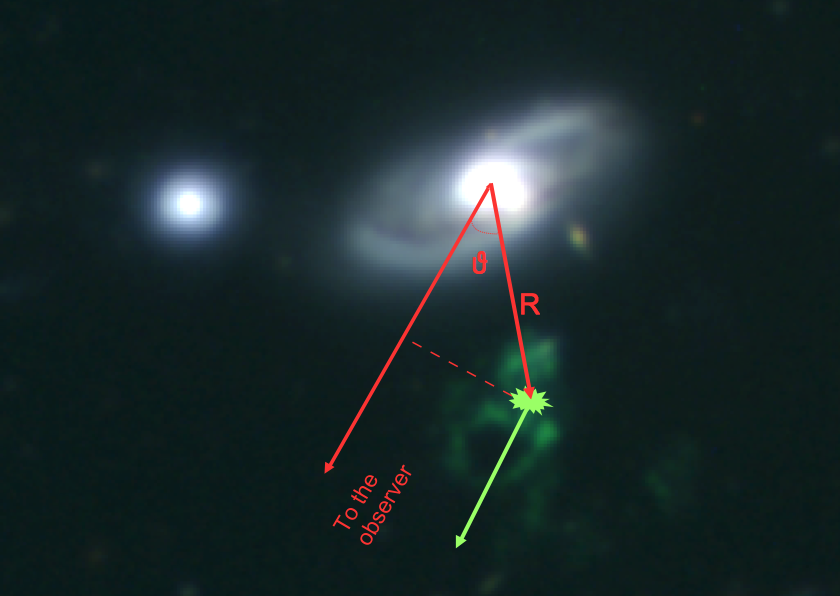}
}
\caption{
Optical image of the galaxy IC~2497 with its companion and the Hanny's Voorwerp nebula {(in green)} obtained by the authors at the 6m telescope in $BVR$ filters in  2011 Oct.  
A geometric scheme   illustrating the eq.(\ref{moiseev:eq1}) for  the signal delay from the nucleus is overlapping. 
}
\label{Moiseev:HV}
\end{figure}

The prototype of the discussed class of objects is Hanny's Voorwerp (translated from Dutch --- `Hanny's Object') --- a nebula discovered   by the  volunteer of the Galaxy Zoo citizen project \citep{Lintott2008MNRAS.389.1179L}  Hanny van Arkel  on SDSS images at the projected distance $15-30$ kpc from   the spiral galaxy IC 2497 (fig.~\ref{Moiseev:HV}, see also \cite{Keel_HV2012AJ....144...66K}). The follow-up spectroscopic observations found   that the nebula is {an objects at the same redshift with IC~2497 and its} spectrum contains strong forbidden lines \OIII, \NII, \SII{} typical for Seyfert galaxies.  The relatively bright lines of helium and neon  indicate a high electron temperature --- at least 10,\,000 --- 20,\,000~K, which could provide either illumination by the AGN or shock waves with velocities  $\sim400\kms$, what  doesn't match to the observed gas kinematics. Meanwhile, the spectrum of the host galaxy indicates only a central star-formation burst and weak LINER-type activity.  The observed luminosity in the infrared range is inconsistent with the assumption of a powerful AGN hidden  by  dust clouds. It was reasonable to assume that the luminosity of the ionizing source had been dramatically decreased by almost two orders of magnitude over the last hundred thousand years \cite{Lintott2009}. The layout in Fig. ~\ref{Moiseev:HV} (based on \cite{Keel2012}) explains why we received information about the ionization of the nebula with a delay relative to the photons coming directly from the nucleus.

The general scenario of the Hanny's Voorwerp (HV) origin\footnote{See the illustration in \url{https://esahubble.org/images/heic1102c/}}  looks as follows. Gravitational interaction with a satellite leads both to the formation of a tidal tail emerging from the disk of the host galaxy, and to the inflow gas motions into the disk that started feeding SMBH and as, a consequence, triggered AGN.  The  nebula appears in emission lines because  the gas in the tidal tail   getting into the ionization cone.  When the gas in the circumnuclear region have been depleted, the relaxation stage occurred and ionization radiation was switched off, but the nebula is still visible for some time for the observer.

From the diagram shown in Fig.~\ref{Moiseev:HV},  the signal delay of the nuclear activity changing is (see \cite{Keel2012}):
\begin{equation}
\label{moiseev:eq1}
  \Delta t =R(1-\cos\theta)/c=\frac{r_{obs}(1-\cos\theta)}{c\sin\theta}\,,
\end{equation}
where $R$ is the true distance from the nebula to the nucleus, $r_{obs}$ is  the observed projection distance, $\theta$ is the angle between the directions to the nebula and to the observer. 

Since $\theta$ is usually unknown,  in the papers cited below  usually assumed  $\Delta t\approx r_{obs}/c$, corresponding to $\theta=90^\circ$. However, as  it follows from the  eq.~(\ref{moiseev:eq1}), true value $\Delta t$  may be very different from this rough estimation in the case of other arrangement of the AGN and  nebula relative to the observer.

\HI~ mapping has confirmed the existence near the IC~2497 the tail-like structure seen in the neutral hydrogen. At the same time, there is a deficit of \HI{} in the region of the HV ionizing nebular, most likely caused not only by ionizing radiation, but also related with a  radio jet kinetic action on the gas \cite{Jozsa2009A&A...500L..33J}. Recent detailed study of the radio jet properties with the LOFAR low frequency radio  telescope gives an estimation of its age  $\sim100$ Myr \cite{Smith2022MNRAS.514.3879S}.  The authors of cited work proposed the following scenario of HV evolution that is more complex than described above: 

\vspace{6 pt}

\fbox{\parbox{0.18\textwidth}
{Tidal encounter left the \HI{} clouds}
}
$\Longrightarrow$
\fbox{\parbox{0.3\textwidth}
{AGN  outburst in a kinetic mode created the radio jet that  punched a hole in the \HI. \\$\sim100$ Myr ago}
}
$\Longrightarrow$
\fbox{\parbox{0.3\textwidth}
{Radiative efficient AGN outburst illuminated and ionized the gas in  HV nebula.  \\~$\sim0.1$ Myr ago}
}


\vspace{6 pt}

In other words, we observe the  signatures of both radio-quiet and radio-loud accretion modes acted during different epoch in the  same galaxy.

The gravitationally perturbed systems with HV phenomenon provide a potential opportunity to calculate a duration of AGN duty cycle as a difference between the time when the radiation outburst started and when the activity stopped according eq.(\ref{moiseev:eq1}). Indeed, it's possible to estimate the age of interaction in the merging pair of galaxies using comparison of observed morphological properties and velocity distribution (velocity field) with results of numerical simulations (the GalMer database is an example of such  library of `snapshots' of galactic collision simulations \cite{Chilingarian2010A&A...518A..61C}).  In this case it  becomes possible to understand the  mutual spatial orientation of the different components (galaxy disk, tidal tails, ionizing nebula) and also recognize the moment of AGN triggering by gas inflow motions related with tidal forces.  
 
Unfortunately, in the   case of IC~2497 there is not enough observational information for this sort of modeling:  stellar tidal structures are absent on the available deep images, whereas  the presented \HI{} map  has too low spatial resolution. There are even doubts that the satellite observed to the east of the main galaxy (Fig.~\ref{Moiseev:HV}) is the intruder of the main galaxy structure. Its luminosity and mass seems to be  too low for  strong tidal disturbance. Therefore, it is important to look for more convenient systems for modeling its spatial structures. Moreover, analysis of the occurrence of HV phenomenon in different sample of galaxies  can help us to better understand the  character of AGN variability at the timescale $10^4-10^5$ years  from simple statistical considerations (see Sec.~\ref{sec:stat}).

\subsection{ \bf New examples: `voorwerpjes'}

\label{sec:newexampl}
How unique is the Hanny's Voorwerp? The first significant attempt  to answer this question was made by \citet{Keel2012} in the survey   based on the visual examination of SDSS images by participants of the Galaxy Zoo project. They  compiled a sample  of all nearby ($z<0.1$) galaxies from SDSS DR7 that were either already in the AGN catalogs or whose line ratio in the SDSS spectra corresponded to possible nuclear activity --- a total of 18116 objects. A close look at the SDSS images revealed 49 candidates with suspected external emission clouds. Follow-up spectroscopy at the 2.1m Kitt Peak National Observatory   and 3m Lick Observatory telescopes confirmed EELR ionized by AGN in  19 galaxies. It is noteworthy that all systems are either clearly interacting \citep[at least 14/19 according][]{Keel2012} or are surrounded by  faint tidal structures \cite{Keel2015}. This fact gives independent  agreement with the Hanny's Voorwerp formation scenario described in the previous section, i.e. that extended ionized gas rises from tidal off-plane structures.

\begin{figure}
\includegraphics[width=0.99\textwidth]{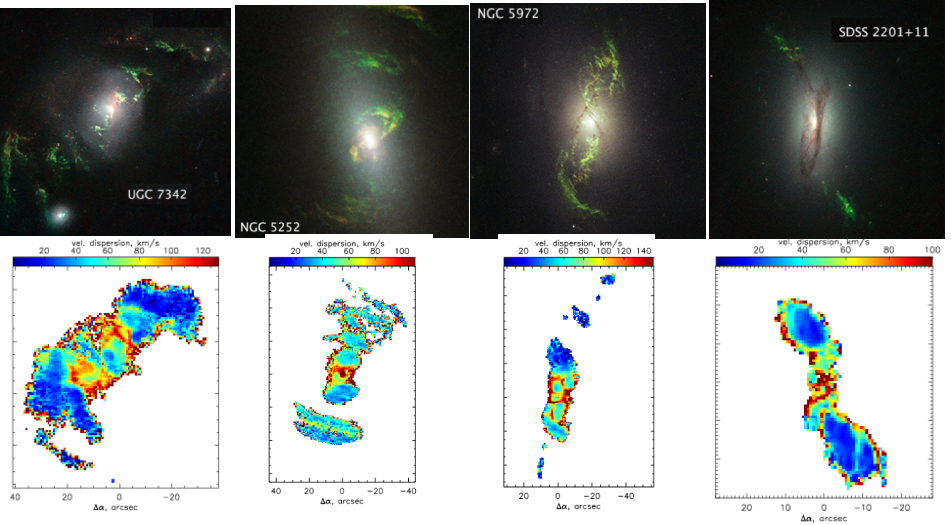}
\caption{Images of Seyfert galaxies with fading nuclear activity obtained with the HST (top), emission in the \OIII{}  is shown in green (credits: NASA / ESA / W. Keel). The bottom panels show the corresponding maps of the ionized gas  velocity dispersion according the  6m telescope observations \cite{Keel2015} (the color scale is in $\kms$). 
}
\label{Moiseev:hst}
\end{figure}
 
An energetic budget of EELRs was estimated via comparison of the total luminosity of the nucleus in the far infrared range, including the contribution from the heated dust ($L_{FIR}$), with the luminosity required for ionization of the detected clouds ($L_{ion}$)   \cite{Keel2015}. It  was found $L_{ion}/L_{FIR}>1$ for 7 of 20 galaxies (including the prototype --- IC 2497)  that indicates a fading  activity. The remaining objects in the sample could be considered as AGN with obscured nuclear UV radiation.

To apply the scenario of the Hanny's Voorwerp formation for the new findings (these were called `voorwerpjes'  \cite{Garrett2012aska.confE..47G}) we have to be sure that the detected gaseous clouds are dynamically  quiet. An alternative explanation of EELRs assumes the gas ejected from nuclear region as a  result of AGN-driven outflows well known in radio-load  quasars \citep{Fu2009ApJ...690..953F, Harrison2014MNRAS.441.3306H,Rupke2011ApJ...729L..27R}. The size of EELR related with such outflows can reach several tens of kpc \citep{Vayner2021ApJ...919..122V,VillarMartin2021A&A...650A..84V}. Also jet-cloud interaction leads to the  strong shock waves producing the emission line ratio similar with the expected for the   ionization cones   described above: an excitation of \OIII, \SII, and  \NII{} forbidden lines together with \HeII$\lambda4686$ emission (the last one, if the shock exceeds 400-500$\kms$, \citep{Keel2012}). 

A detailed investigation of the gas kinematics, i.e. distribution of line--of-sight velocity and velocity dispersion, allows  to  distinguish these two types of EELR  (ionization cone vs AGN outflow). This work was made for the fading AGN candidates in the sample  from \cite{Keel2012}  using 3D spectroscopy with the scanning FPI at the 6m telescope and GMOS integral-field spectrograph \citep{Keel2015, Keel2017ApJ...835..256K}.  It was shown, that  ionized gas in the circumnuclear regions ($r<1-2$ kpc) is  usually   perturbed: it has  significant non-circular motions, the velocity dispersion  in the \OIII{} is also large  ($\sigma\approx100-200\kms$), the emission line profile often has a multicomponent structure. These kinematic features  indicate the dynamic effect of the AGN jet and/or wind outflow at the surrounding  gas. At the same time, at larger distances from the nucleus, the EELR gas is  dynamically cold ($\sigma\approx10-50\kms$),  the velocity field follows a regular rotation. Moreover, as in the outer filaments of Mrk~6  considered  in Sec.~\ref{sec:Mrk6},  the circular rotation does not imply motion in the plane of the host galaxy stellar disk. 

The linear size of ionized gas structures in some voorwerpjes  exceeds significantly their   hosts size. For instance, the nuclear radiation in  Teacup  (SDSSJ 1430+1339) was traced in the surrounding gaseous nebulae up to projected distance 55 kpc \cite{Villar2018MNRAS.474.2302V, Moiseev2023Univ....9...66M}.  The diameter of EELR in NGC  5972 is about 70 kpc \cite{Keel2015}.

In many cases the gas  illuminated by AGN comes from the external environment with the spin direction different  from that of the AGN host. Inclined orbits  must  precess and fall  to the main galactic plane, we can see the complex dust distribution related with shock waves in the  warped and precessed gas in  galaxies NGC 5972 and SDSS~2201+11 HST  images (Fig.~\ref{Moiseev:hst}). The corresponded  dynamical models are given in \cite{Keel2015}. The similar interaction between gas clouds on  misalignment orbits appears in  some gas-rich polar ring galaxies \citep[see][and references therein]{Moiseev2014ASPC..486...61M}. 

Finally, 3D spectroscopic kinematic mapping gives evidences that ENLRs in  found  voorwerpjes  is  ionized by AGN radiation, rather than by shocks. Using the recombination-balance approach to analyse the \Ha{} surface brightness of EELR in  HST images \citet{Keel2017ApJ...835..256K}   traced the nuclear ionization history   in 8 candidates to the fading AGN selected as described above. It was found that  the brightness of regions further from the nucleus corresponds to a higher bolometric luminosity of the nucleus than currently observed. Sometimes these changes are smooth, sometimes its appear  rapid drops. But everywhere in the considered sample the   luminosity of the central source has been decreasing significantly in the last $\sim20\,000$ years.  

{The lower limit of the $\Delta t$ estimation is related with the ionized gas recombination time-scale. The equation (\ref{moiseev:eq1}) is written for a reflecting nebula, whereas the typical recombination time differs for various combinations of temperature and electron  density $n_e$ and   
can be roughly estimated as \citep{Osterbrock2006agna.book.....O}:
\begin{equation}
\label{moiseev:eq2}
  \tau_r \approx (a(ion,T)n_e)^{-1},
\end{equation}
where $a(ion,T)$ is the total recombination coefficient of the considered ion. According \cite{Osterbrock2006agna.book.....O} for the ionized hydrogen and $T=10,\,000$~K  in the usually accepted `Case B  approximation': $a(\mbox{H}^{+})=2.6\cdot10^{-13}$~cm$^3$c$^{-1}$. That gives  $\tau_r=1,\,200-12,\,000$ years for the density range   typical in external EELRs \citep[$n_e=10-100$~cm$^{-3}$][]{Keel2012,French2023ApJ...950..153F,Moiseev2023Univ....9...66M}. It means that in the case of low density gas the  observed response in the Balmer emission lines (\Ha,\Hb, etc.) in some considered objects is more smoothed comparing with the pure reflection case in (\ref{moiseev:eq1}). However, high ionized ions recombine significantly faster than $\mbox{H}^{+}$. For instance, the \OIII{} emission disappears  during the time $\tau_r=20-200$ years in the same density range for $a(\mbox{O}^{2+})=1.7\cdot10^{-11}$~cm$^3$c$^{-1}$ \citep{Osterbrock2006agna.book.....O,Crenshaw2010AJ....139..871C}. Moreover, in the real plasma  the charge  exchange  reactions make $\tau_r$ even shorter for heavy elements ions \citep{Binette1987A&A...177...11B}. It means that the recombination time  is not significantly affected the study of  EELRs  selected according their high   \OIII$/\mbox{H}\beta$ flux ratio (see the next section).
}

\section{EELR Statistical Study}
\label{sec:stat}
\subsection{Surveys  of nearby AGNs}

The study of individual unique extended ionization cones shades light into the history of AGN triggering and fading  of specific galaxies. The smaller and fainter analogue of HV was found in the post-merger galaxy NGC 7252 \citep{Schweizer2013ApJ...773..148S}, whereas CFHT and VLT observations of the  significantly more distant ($z=0.326$) galaxy J2240-0927  reveal  ENLR  $\sim70$ kpc in size \citep{Schirmer2013ApJ...763...60S}. This system of  ionized gas clouds has  \OIII{} luminosity on two order larger than in HV and it is also considered as a result of AGN ionization echo. An EELR with  properties  resemble other voorwerpjes  was found  in the post-starburst galaxy PGC 043234 \cite{Prieto2016ApJ...830L..32P} with the MUSE integral-field spectrograph. A very interesting EELR was found in the multi-wavelength study of the interacting pair of galaxies SDSS J1354+1327 \citep{Comerford2017ApJ...849..102C}. The authors argued that the $\sim10$ kpc ionization cone is a result of previous AGN activity, whereas  after $<0.1$ Myr a new AGN outflow was launched.

The extensive search of EELR candidates by volunteers of the Galaxy Zoo project continues under supervision by Prof. William Keel (University of Alabama) including follow-up snapshot imaging at the HST in the frameworks of  Gems of the Galaxy Zoos  project \citep{Keel2022AJ....163..150K}. In addition to SDSS, new GalZoo search is based on deeper large-field  digital sky surveys like DESI Legacy Imaging Surveys\footnote{\url{https://www.legacysurvey.org/}}. A follow-up spectral confirmation of new candidates is carried out at the SAO RAS 6m telescope with SCOPRIO-1/2 multi-mode spectrographs. 

The special cases listed above  has motivated the search of EELR and the fading AGN phenomenon in  samples of galaxies  constrained by various criteria in order to  minimize the observational bias. 3D spectroscopic data and medium-band filters centered on the doublet \OIII{} possesses a significant advantage over broad-band sky image surveys  in the search of a low contrast nebular emission.  

The approach based on an integral-field massive survey was recently illustrated by  \citet{French2023ApJ...950..153F}. The authors gave evidences of 6 EELRs in the sample of 93 post-starburst galaxies selected in Mapping Nearby Galaxies at APO survey \citep[MaNGA][]{MANGA2016AJ....151....8Y}. It was shown that  5/6 of found EELRs has ionized by AGN with fading luminosity. The full duty cycle of AGN in   post-starburst systems was estimated as $\sim0.1$ Myr whereas the luminous phase  continues during only about 5\% of this time. However the typical radius of these  \OIII{} nebulae don't exceed 10 kpc, that is smaller than in HV and many other systems considered in our review. 

The second approach based on the medium-band filters survey is developed in the project   TELPERION (Tracking Emission Lines to Probe Extended Regions Ionized by Once-active Nuclei \cite{Knese2020MNRAS.496.1035K, Keel2022MNRAS.510.4608K}). The search of candidates   was performed with 1m SARA  robotic telescopes with a filter centered on the doublet \OIII$\lambda\lambda4959,5007$ for galaxies spanning  redshifts $z=0.009-0.029$. The detection limit of extended emission  was about 1/10 of the Hanny's Voorwerp surface brightness. 

\begin{figure}
\includegraphics[width=0.99\textwidth]{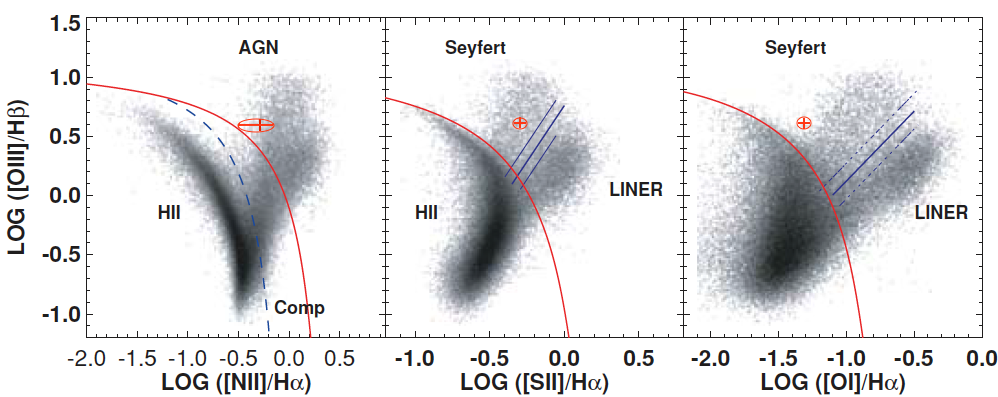}
\caption{Diagnostic BPT-diagrams from \cite{Kewley2006MNRAS.372..961K}, with the flux ratios  in the forbidden oxygen, nitrogen, and sulfur lines to  the hydrogen Balmer lines. The pairs of lines are close in wavelength,   hence the corresponded  ratios is not  affected by internal extinction.  The bounding lines separate the areas corresponding to ionization by young hot stars (HII), the UV continuum of the active nucleus (AGN, Seyfert), and the weak AGN a prominent  contribution of shocks (LINER). The positions of galaxy nuclei from the SDSS survey are shown in gray, forming a characteristic `seagull diagram', where the left wing is the sequence of star-forming regions and the right wing is the active galactic nuclei sequence. The red circle with a cross shows the  EELR in NGC 5514 (see the following figure). 
}
\label{Moiseev:BPT}
\end{figure}

Strictly speaking,  not only ionization cones, but also other type of  sources could produce  \OIII{} emission in galaxy outskirt: \HII{} regions, galactic wind outflow driven by central star burst, etc. Optical spectroscopy of selected candidates  allows to separate gas clouds with different mechanisms of ionization on the diagnostic diagrams plotting the characteristic flux ratios in the pair of emission lines close in wavelength (\OIII$/\mbox{H}\beta$, \NII$/$\Ha{}, etc.), called `BPT-diagrams' after the first authors of the paper \cite{Baldwin1981PASP...93....5B}. Different areas on these  diagrams correspond to the gas ionized by UB-radiation of OB-stars in \HII-regions, by different types of AGN (UV-radiation in Seyfert galaxies and shocks in LINERs) and by    `composite' ionization  between  AGN and HII areas (Fig.~\ref{Moiseev:BPT}).  Such diagrams are often supplemented by grids of model calculations of photoionization or shock excitation, taking into account metallicity, shock velocity, magnetic field, etc. \cite{ITERA2010NewA...15..614G,Postnikova2023AstL...49..151P}.

Follow-up spectroscopy of the TELPERION, candidates were conducted with several  Russian telescopes equipped with high-efficient low-resolution spectrographs developed in  SAO RAS under the supervision of Prof. Victor Afanasiev: SCORPIO-1 and SCORPIO-2 prime focal reducers   at the SAO RAS 6m telescope
spectrograph, ADAM at the  1.6m telescope AZT33-IK of Sayan Observatory\footnote{The description of all listed devices can be found at this  URL:\url{https://www.sao.ru/hq/lsfvo/devices_eng.html}}.  For the most interesting confirmed EELRs, deep \OIII{} images were obtained  at the 6m telescope and at the 2.5m telescope of the SAI MSU using the tunable filter photometer MaNGaL (Mapper of Narrow Galaxy Lines\footnote{\url{https://www.sao.ru/hq/lsfvo/devices/mangal}}). The MaNGaL instrument uses scanning FPI as a narrow-band filter, about 1.3~nm wide, whose transmission can be fine-tuned to the wavelengths corresponding to both the emission line  and the continuum, taking into account the redshift of a target galaxy. 

With the above technique the  first search for distant emission clouds in a luminosity-limited sample of nearby AGN was done \cite{Keel2022MNRAS.510.4608K}. The sample contained 111 galaxies --- all known AGN  brighter than $M=-20^m$ in the redshift range of $z=0.009-0.029$ and $\delta>-36^\circ$. Among 15  emission clouds discovered  in \OIII{} images, 6 gaseous systems were  confirmed as EELRs via follow-up spectroscopy. Two of them (in NGC~235 and NGC~5514) are  distant clouds projected at $r=25-75$ kpc from the nucleus.

\begin{figure}
\includegraphics[width=0.99\textwidth]{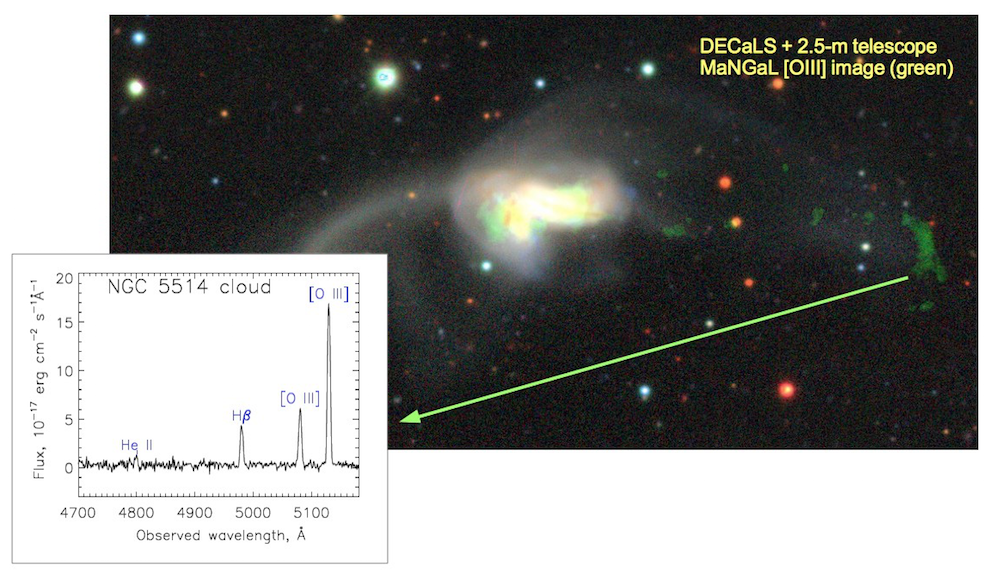}
\caption{
Optical image of the interacting galaxy system NGC~5514 from the DESI Legacy survey \citep{Legacy2019AJ....157..168D}. The green color shows the emission distribution in the ionized oxygen line according to the observations with the MaNGaL instrument at the 2.5m SAI MSU telescope. The gas illuminated by the AGN is visible far beyond the disks of galaxies inside the tidal arm. The box shows the spectrum of clouds obtained with the 6m SAO RAS telescope using the SCORPIO-2 instrument. The lines of the ionized oxygen and helium are visible confirming the gas ionization by the active nucleus \cite{Keel2022MNRAS.510.4608K}.
}
\label{Moiseev:N5514}
\end{figure}

NGC~5514 (Fig.~\ref{Moiseev:N5514}) exhibits  an almost classical illustration of  the described above scenario of the Hanny’s Voorwerp formation. In merging pair of galaxies  the interaction leads to the gas inflow to the central SMBH in one of the galaxies. The \OIII{} emission reveals   both an inner ionization cone ($r<12$ kpc) and more distant gas clouds in the tidal tail up to 75 kpc away in projection distance. 
Whereas in the circumnuclear region  signs of AGN-induced outflow is observed, the EELR gas is kinematically quiescent, i.e. it is characterized the low gas velocity dispersion without significant non-circular gas motions. Also the detection \HeII$\lambda4861$ emission line (see the spectrum in Fig.~\ref{Moiseev:N5514}) clearly indicates the ionization by hot source like AGN accretion disk. An alternative explanation is a strong  shock excitation, but it   the corresponded shock velocities ($>400\kms$)   disagree with a dynamically cold gas.
Also an emission-line ratios indicate the AGN-type photoionization of the distant clouds in NGC 5514 (Fig.~\ref{Moiseev:BPT}). An estimation  of the energetic budget of AGN using the $L_{ion}/L_{FIR}$ criterion shows that activity in NGC~5514 has decreased by more than a factor of 3 in the last $\sim0.25$~Myr \cite{Keel2022MNRAS.510.4608K}. 

\begin{figure}
\includegraphics[width=0.99\textwidth]{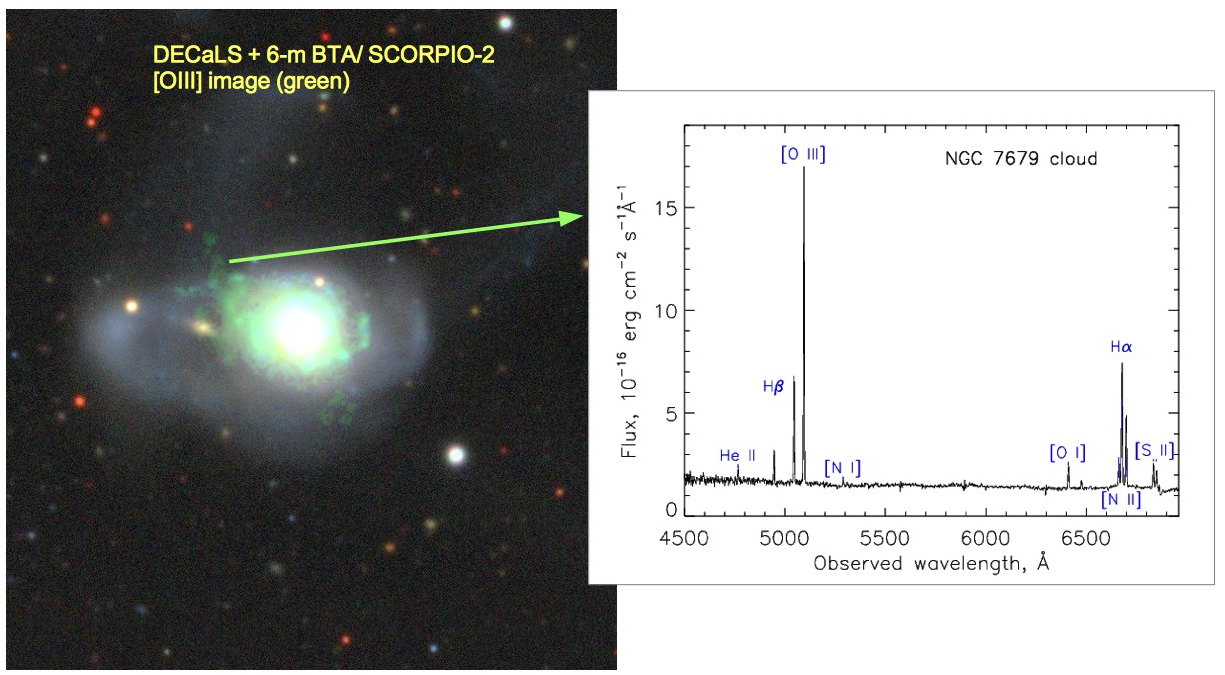}
\caption{
The same as Fig.~\ref{Moiseev:N5514} for  the galaxy NGC 7679 whereas the \OIII{} image was conducted with the SCORPIO-2 at the SAO RAS 6m telescope. The AGN illuminates the gas inside the galaxy \cite{Keel2022MNRAS.510.4608K}.
}
\label{Moiseev:N7679}
\end{figure}

In addition to NGC~235  and NGC~5514, large (up to 10 kpc in size) ionization cones inside galactic disks have been found in three  galaxies: ESO~362-G08, NGC~7679 (Fig.~\ref{Moiseev:N7679}) and IC~1481. Interestingly, the latter, as well as NGC~5514, contains a low activity AGN --- LINER. Why nucleus with a relatively low ionizing flux produces such prominent cones is not yet very clear. Perhaps this is another indication of the transient activity phenomenon of galactic nuclei in combination with properties of surrounding interstellar medium (gas distribution, its density, metallicity, etc.). Interesting that the prototype voorwerpjes galaxy --  IC~2497 (the Hanny’s Voorwerp) is also classified as a LINER. 

Despite the fact that the statistics is not yet very large,  the  EELRs outside the galaxy stellar disks are detected  among 2-5\% percent of AGNs depending on selection of samples: bright AGN and Toomre merging sequence \cite{Keel2022MNRAS.510.4608K}, AGN with  \HI{} external structures \cite{Knese2020MNRAS.496.1035K} or their combinations. Recent counting of the total TELPERION sample of 241 galaxies gives the EELRs incidence  $1.7\pm0.6$\% \cite{Keel2024}. This fraction in several tens times higher that in the previous Galaxy Zoo survey based on SDSS broad-band images with higher surface brightness threshold.  Moreover  the EELR detection rate  is significantly higher in   AGN hosted in interacting and merging galaxies ($10-12\%$, \cite{Keel2022MNRAS.510.4608K, Keel2024}). The off-plane gas in tidal tails  enters the cone of illuminating `spotlight'  with higher probability and on the larger distances than gas in a galaxy plane.

\subsection{Cross-ionization of a companion}

In all  cases discussed above, we learned about a  `screen' (a system of gas clouds) because it got inside the AGN ionization. However we can look on the same problem  in a different   angle:  try to determine  the radiation cone parameters if the  screen geometry is already known.
 
 \begin{figure}
\includegraphics[width=0.99\textwidth]{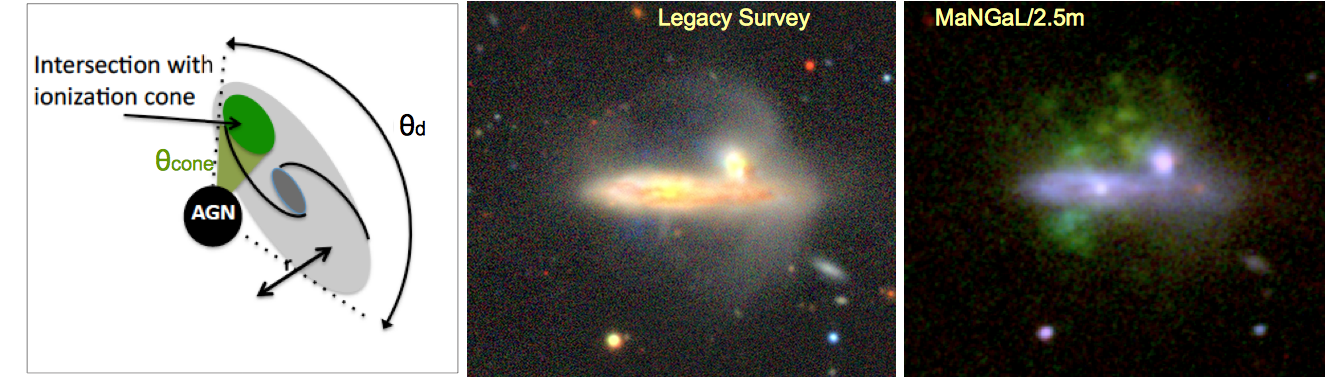}
\caption{Cross-ionization in galaxy pairs based on \cite{Keel2019MNRAS.483.4847K}. From left to right:  the cartoon explaining the geometry of the system consisting of a pair with AGN and  nearby spiral galaxy, whose disk is observed from the nucleus at the angle of $\theta_d$;  the image of pair of galaxies UGC~6081 according the DESI Legacy survey;  the same field was mapped with the  MaNGaL instrument on the SAI MSU 2.5m telescope (emission in the \OIII,    \Ha{} and continuum in the green, red and blue channels correspondently).  
}
\label{Moiseev:cross}
\end{figure}

The first example of known geometry of the screen is a pair   where at least one of the companion galaxies contains AGN and the other one is a gas-rich disk galaxy. In this case we can expect cross-ionization effect, when an EELR is created  in the companion galaxy's disk. Such `ionizing footprint from a neighbour's  even allow to find hidden AGN in an interacting system, as it was shown in Was~49 \citep{Moran1992AJ....104..990M}. An inverse case is  ShaSS~073 in  the Shapley supercluster, where the EELR excited in the companion's disk gave evidences of a  dramatic fading of AGN  luminosity (in 20 times in the past 30\,000 yr) hosted in the  second component of the pair at projected separation 21 kpc \citep{Merluzzi2018ApJ...852..113M}. {The most extreme example of cross-ionization among nearby galaxies was recently found in Mrk~783. The AGN in the main  galaxy ionizes not only the surrounding tidal structure up to the distance 41 kpc, but also the external part of the disk of the companion galaxy SDSS~J130257.20+162537.1 at the projected distance 99~kpc \citep{Moiseev_2023Mrk783}. The preliminary calculation of the ionized balance  indicates that there is no significant decreasing ionizing radiation during last 0.1–0.3 Myr. New in-depth studies of this and similar pairs with  integral-field facilities on large telescopes will help to better restore the history of AGN radiation output.}

In the close pairs of galaxies it is possible to  estimate  the solid angle at which the companion disk is visible from the nucleus (see the diagram in Fig.~\ref{Moiseev:cross}). Therefore, the width of the ionization cone can be constrained from  the statistics of the  EELR occurrence in AGN pairs. It was done in the paper \cite{Keel2019MNRAS.483.4847K} considered the sample of pair with Seyfert nucleus in one or both companions selected in the SDSS DR8 according the following criteria: the  distance between the nuclei $r<15''$($\sim18$ kpc for the median redshift $\langle z\rangle=0.06$), difference in radial velocities within $400\kms$  implies a gravitational bound pairs.  Out of 212 pairs, 32 were selected for spectral observations, distinguished by the smallest distance between centers and the largest $\theta_{d}$ angle.  Fig. ~\ref{Moiseev:cross} shows one of the most interesting  system -- UGC~6081. The nuclei of both companions are active and  two systems of ionization cones are observed simultaneously. 

Given the detection rate of cross-ionization occurrence  (10/32), the average width of the ionization cone in the sample was evaluated  as $\theta_{cone}\approx70^\circ$ in good agreement with earlier estimations by other methods.

A different value of  $\theta_{cone}$ was obtained for the other type of a screen with known geometry: in the sample of AGN with  extended systems of neutral hydrogen \cite{Knese2020MNRAS.496.1035K}  mapped in 21 cm line.  From  26 \HI-rich systems only one EELR was found. It was the gas clouds projected in 12 kpc away from the nucleus of Seyfert galaxy Mrk~1 that has a common \HI{} envelope  with NGC~451 \cite{Knese2020MNRAS.496.1035K}. Calculation gives $\theta_{cone}<20^\circ$ if the AGN  are continuously bright for scales longer than the light-travel times to the external \HI{} structures. The contradiction with previous estimation from cross-ionized pairs is explained if the  AGN ionizing radiation  undergoes strong variations  at times of $10^4-10^5$ years. This characteristic time is similar that was presented  above  for fading AGN.

The examples considered in this review are only the first attempts to estimate the characteristics of the AGN accretion machine from the statistics of the occurrence of its ionization trace. Increasing of the samples of such objects as well as more detailed study will make it possible to answer a number of relevant and interesting questions in AGN physics, related, for example, with the precession of the ionization cone or with evolution of  dust torus collimating the AGN UV radiation. For this purpose, it is relevant to carry out a deep mapping of the environment of nearby  galaxies in the main emission lines (\Ha, \OIII, etc.).
Even the first attempts in this direction are already yielding interesting results. Deep   \Ha{} imaging of famous nearby   interacting pair  `The Whirlpool' (NGC 5194/5) revealed an extended low-brightness cloud at 32 kpc away from the center of the main galaxy. The properties  of gas excitation correspond  to that observed in classical  ionization cones  \cite{Watkins2018ApJ...858L..16W, Xu2023ApJ...943...28X}. 

It is likely  that further study of the Local Group galaxies, will allow to detect new traces of the ionization cones. Accordingly, the question arises: what is about our Milky Way?

\section{Activity of the Milky Way nucleus}

Indeed, the SMBH is nesting in the Milky Way nucleus.  This  fact was confirmed both directly by images \cite{EHT2022ApJ...930L..12E} and by mass measurements from the kinematics of nearby stars (the Nobel Prize was awarded in 2020 to Reinhard Genzel and Andrea Ghez `for the discovery of a supermassive compact object at the centre of our galaxy'). The mass of the SMBH ($4.15\pm0.01\cdot10^6\,M_\odot$, \cite{SgrAmass2019A&A...625L..10G}) is not very large compared to powerful quasars, but quite sufficient for noticeable manifestations of activity.    
Such manifestations are known. First of all, its are `Fermi bubbles', named after the Fermi Gamma-ray Space Telescope, which discovered in 2010 two giant spherical structures emitting in the gamma-ray, extending up to $\pm10$ kpc perpendicular to the galactic plane from the Milky Way center \cite{Fermi2010ApJ...724.1044S}. Formally say, this morphology can also be explained by the galactic wind driven by a circumnuclear burst of star formation. However the required rate of supernovae explosion is much higher than expected. At the same time, the observed specific magnetic field distribution and energy spectrum of cosmic particles of Fermi bubbles are explained in models of recent activity of the Milky Way's central black hole (Sgr A$^*$), suggesting the age of the bubbles  about 2 Myr and a total energy more than $10^{55}$ erg \cite{Barkov2014A&A...565A..65B}. 

The intrigue intensifies when a system of X-ray eROSITA  bubbles \cite{Predehl2020Natur.588..227P}  named after the eROSITA telescope of the Russian-German space observatory `Spectrum-R\"oentgen-Gamma' was found. These structures of hot gas emitting in the soft X-ray are larger (about 14 kpc) and  can be consedered as extended envelope of Fermi bubbles. The whole set of observations of both space telescopes can be explained by assuming that it is a single act of energy release in the center of the Galaxy. In this case, the Fermi bubbles correspond to the inner boundary of the most heated interstellar medium, while the eROSITA bubbles correspond to the propagation of the expanding shock wave in the Galactic gaseous halo. The total energy of the process is estimated  $\sim10^{56}$ erg, and the age is  1-2 Myr \cite{Predehl2020Natur.588..227P}. Recently a detailed numerical model has been presented suggesting that it was the activity of the Sgr~A$^*$ jet that formed the observed hot gas bubble system \cite{FermieROSITA2022NatAs...6..584Y}. 

What is about the ionizing activity of the Milky Way's nucleus? 
In addition to the gas in the galactic disk the system of  outer clouds is exist, that allows us to detect a possible trace of the `ionization spotlight'. It is the  Magellanic Stream --- a giant arch of neutral hydrogen  stretching almost through the entire southern galactic hemisphere. There is virtually no doubt that this gas is associated with satellites of the Milky Way --- the Large and Small Magellanic Clouds. Most likely it's a  matter losted by them under the action of  the Galaxy gravity. \citet{Bland-Hawthorn2013ApJ...778...58B,Bland-Hawthorn2019ApJ...886...45B} showed that some part of the gas in Magellanic Stream is ionized, with the  ionization properties corresponded to the excitation by   hard radiation from the Seyfert nucleus.  

The axis of the inferred ionization cone deviates only $15^\circ$ from the Galactic  rotation axis, and the width of the cone ($\sim60^\circ$) agrees well with the estimates given above for such structures in other AGN. The time elapsed since the ionization activity  ($3.5\pm0.1$ Myr) is quite consistent with the age of the Fermi-eROSITA bubbles, therefore it is likely to be a single act of activity. Any case, new breakthroughs in both observations and numerical simulations should be expected in this field of Galaxy research. 

\section{Ionization cones and nuclear outflows}
\label{sec:outflow}

{Our  review is mostly focuses on the search and study  of kinematically-quiet EELRs  that  are in dynamical equilibrium with the host galaxy, which can be considered as one of the best tracer both of the AGN ionization output and low-density intergalactic medium. However, as it was already mentioned in Sec.~\ref{sec:newexampl}, there is a population of kinematically hot EELRs produces by  nuclear outflows driven by jet-cloud interaction or thermal energy and radiation pressure from the accretion disk \citep[see the recent review][and references therin]{Singha2023Galax..11...85S}. A detailed description of this phenomenon is  beyond the  scopes of this review, however in many cases we observe a mix of both acts, i.e. kpc-scale outflow  ionized by AGN. One of the nearest luminous  quasar --- the Teacup galaxy ($z=0.09$) exhibits this kind of composite structure. A well-described  $\sim12$ kpc ionized gas bubble  inflated by the AGN outflow was included in the list of voorwerpjes with ionization balance corresponds to fading AGN \citep{Keel2015}. Long-slit spectroscopy at the Gran Telescopio Canarias revealed that this bubble is only internal part of a giant emission nebula  $\sim100$ kpc in size \citep{Villar2018MNRAS.474.2302V}. The nebula was mapped in the \OIII{} emission line at the 6m telescope, it was shown that the external emission arcs might be a remnant of the previous outflow (the age $<0.8$ Gyr) ionized by the AGN radiation \citep{Moiseev2023Univ....9...66M}.  Whereas the forthcoming integral-field observations with MUSE at the 8m Very Large Telescope allows to study in details    properties of the ionized gas outflow up to 30 kpc from the AGN \citep{Venturi2023A&A...678A.127V}
}

{Teacup AGN can be considered as a local prototype of combined influence of the AGN radiation and outflow on the surrounding matter on scales of the host galaxy and beyond in the high redshift Universe. These effects were considered in set of cosmological simulations \citep[for instance,][]{Bieri2017MNRAS.464.1854B,Kakiichi2017MNRAS.468.3718K,Graziani2018MNRAS.479.4320G}. It is possible that \CII{}-emitted haloes  recently discovered in star forming and AGN early galaxies at $z\sim6$ \citep{Cicone2015A&A...574A..14C,Fujimoto2019ApJ...887..107F,Pizzati2020MNRAS.495..160P} are evolutionary related with  local EELRs considered in our review.  New space telescope JWST is ushering in a new era of study AGN outflows and ionization cones in `cosmic noon' period ($z=1-3$) as it was already demonstrated in the first published results \citep{Vayner2023ApJ...955...92V,Veilleux2023ApJ...953...56V}.
}

\section{Conclusion and  Future Directions}

The authors hope that in this brief review they have been able to show how  observations of extended (and distant) emission clouds provide information about the kinematics and state of gas ionization at distances up to tens and even hundreds of kpc from the center of galaxies. The study of  the active nucleus ionizing  cones makes it possible to trace the processes of low-density cold gas accretion  by galaxies with much better angular resolution than in \HI{}  to examine the activity history of a supermassive black hole on time scales of $10^4$ -- $10^5$ years, and even to constrain  the parameters of the accretion machine hidden inside the central parsec. Moreover, such studies are also relevant in the case of our own Galaxy. 

\vspace{6pt} 


\authorcontributions{Conceptualization, A.M. and A.S.; writing---original draft preparation, A.M.; writing---review and editing, A.S. All authors have read and agreed to the published version of the manuscript.
}

\funding{ 
The work was performed as part of the SAO RAS government contract approved by the Ministry of Science and Higher Education of the Russian Federation.
}


\acknowledgments{This work is dedicated to the memory  of  Victor   Afanasiev  whose  developed the spectral instruments for 6m and 1.6m telescopes  and  to the memory of Victor  Kornilov whose enthusiasm and knowledge make the 2.5m telescope work successfully.
We thank the reviewers for their constructive comments. Observations with the SAO RAS telescopes are supported by the Ministry of Science and Higher Education of the Russian Federation. The renovation of telescope equipment is currently provided within the national project "Science and Universities". 

The Legacy Surveys consist of three individual and complementary projects: the Dark Energy Camera Legacy Survey (DECaLS; Proposal ID \#2014B-0404; PIs: David Schlegel and Arjun Dey), the Beijing-Arizona Sky Survey (BASS; NOAO Prop. ID \#2015A-0801; PIs: Zhou Xu and Xiaohui Fan), and the Mayall z-band Legacy Survey (MzLS; Prop. ID \#2016A-0453; PI: Arjun Dey). DECaLS, BASS, and MzLS together include data obtained, respectively, at the Blanco telescope, Cerro Tololo Inter-American Observatory, NSF’s NOIRLab; the Bok telescope, Steward Observatory, University of Arizona; and the Mayall telescope, Kitt Peak National Observatory, NOIRLab. NOIRLab is operated by the Association of Universities for Research in Astronomy (AURA) under a cooperative agreement with the National Science Foundation.

This project used data obtained with the Dark Energy Camera (DECam), which was constructed by the Dark Energy Survey (DES) collaboration. Funding for the DES Projects has been provided by the U.S. Department of Energy, the U.S. National Science Foundation, the Ministry of Science and Education of Spain, the Science and Technology Facilities Council of the United Kingdom, the Higher Education Funding Council for England, the National Center for Supercomputing Applications at the University of Illinois at Urbana-Champaign, the Kavli Institute of Cosmological Physics at the University of Chicago, Center for Cosmology and Astro-Particle Physics at the Ohio State University, the Mitchell Institute for Fundamental Physics and Astronomy at Texas A\&M University, Financiadora de Estudos e Projetos, Fundacao Carlos Chagas Filho de Amparo, Financiadora de Estudos e Projetos, Fundacao Carlos Chagas Filho de Amparo a Pesquisa do Estado do Rio de Janeiro, Conselho Nacional de Desenvolvimento Cientifico e Tecnologico and the Ministerio da Ciencia, Tecnologia e Inovacao, the Deutsche Forschungsgemeinschaft and the Collaborating Institutions in the Dark Energy Survey. The Collaborating Institutions are Argonne National Laboratory, the University of California at Santa Cruz, the University of Cambridge, Centro de Investigaciones Energeticas, Medioambientales y Tecnologicas-Madrid, the University of Chicago, University College London, the DES-Brazil Consortium, the University of Edinburgh, the Eidgenossische Technische Hochschule (ETH) Zurich, Fermi National Accelerator Laboratory, the University of Illinois at Urbana-Champaign, the Institut de Ciencies de l’Espai (IEEC/CSIC), the Institut de Fisica de Altes Energies, Lawrence Berkeley National Laboratory, the Ludwig Maximilians Universitat Munchen and the associated Excellence Cluster Universe, the University of Michigan, NSF’s NOIRLab, the University of Nottingham, the Ohio State University, the University of Pennsylvania, the University of Portsmouth, SLAC National Accelerator Laboratory, Stanford University, the University of Sussex, and Texas A\&M University.

BASS is a key project of the Telescope Access Program (TAP), which has been funded by the National Astronomical Observatories of China, the Chinese Academy of Sciences (the Strategic Priority Research Program ‘The Emergence of Cosmological Structures’ Grant \#XDB09000000), and the Special Fund for Astronomy from the Ministry of Finance. The BASS is also supported by the External Cooperation Program of Chinese Academy of Sciences (Grant \# 114A11KYSB20160057), and Chinese National Natural Science Foundation (Grant \# 11433005).

The Legacy Survey team makes use of data products from the Near-Earth Object Wide-field Infrared Survey Explorer (NEOWISE), which is a project of the Jet Propulsion Laboratory/California Institute of Technology. NEOWISE is funded by the National Aeronautics and Space Administration.

The Legacy Surveys imaging of the DESI footprint is supported by the Director, Office of Science, Office of High Energy Physics of the U.S. Department of Energy under Contract No. DE-AC02-05CH1123; by the National Energy Research Scientific Computing Center, a DOE Office of Science User Facility under the same contract; and by the U.S. National Science Foundation, Division of Astronomical Sciences under Contract No. AST-0950945 to NOAO.}

\conflictsofinterest{The authors declare no conflict of interest.} 

\abbreviations{Abbreviations}{
The following abbreviations are used in this manuscript:\\
\noindent 
\begin{tabular}{@{}ll}
AGN  & Active galaxy nucleus\\
EELR & Extended Emission-Line Regions \\
FPI & Fabry-Perot interferometer\\
JWST & James Webb Space Telescope \\
HST & Hubble Space Telescope \\
HV &   Hanni's Voorwerp \\
LINER & Low-ionization nuclear emission-line region \\
LOFAR &  International Low-Frequency Array \\
MUSE &  Multi Unit Spectroscopic Explorer \\
SAI MSU & Sternberg Astronomical Institute of Lomonosov Moscow State University  \\
SAO RAS & Special Astrophysical Observatory of the Russian Academy of Sciences \\
SCORPIO & Spectral Camera with Optical Reducer for Photometric and Interferometric Observations \\
SDSS & Sloan Digital Sky Survey \\
Sy & Seyfert galaxy\\
\end{tabular}
}

\begin{adjustwidth}{-\extralength}{0cm}

\reftitle{References}

\end{adjustwidth}
\end{document}